\documentclass{astlb} 
\usepackage[english]{babel}
\usepackage[utf8]{inputenc} 
\usepackage{amsmath}
\usepackage{amssymb}                                          
\usepackage{graphicx}

\usepackage{natbib}
\setcitestyle{numbers, open={[[}, close={]]}}

\usepackage{mathtext}
\usepackage[hyperfootnotes=false]{hyperref}
\usepackage{color}
\usepackage{epsf}
\usepackage{amsfonts}
\usepackage{gensymb}
\usepackage{amssymb}
\usepackage{float}
\usepackage{ulem}
\usepackage{diagbox}

\def\tightleading{1.1}
\def\doubleleading{1.6}
\def\baselinestretch{\doubleleading}

\let\tightenlines=\tighten

\tightenlines

\begin{document}
%\baselineskip 21pt

%\journalinfo{2017}{0}{0}{1}[0]
%\UDK{524.77}

\title{Annihilation of positrons from AGN jets as a possible source of cosmic gamma-ray background at energies below 511~keV}
\author{B.A. Nizamov$^{1}$\email{nizamov@physics.msu.ru}, M.S. Pshirkov$^{1,2,3}$,   \\
$^1$ Sternberg Astronomical Institute, Lomonosov Moscow State University, Universitetsky pr., 13, Moscow, 119234, Russia\\
$^2$ Faculty of Physics, Lomonosov Moscow State University, \\ Leninskie Gory 1-2, 119991 Moscow, Russia\\
$^3$ Lebedev Physical Institute, Pushchino Radio Astronomy Observatory\\
}

\submitted{00.00.2018}

\pagebreak
\begin{abstract}
The origin of the diffuse gamma-ray background in the range from hundreds keV to several MeV is not known conclusively. From current models and observations it is believed that, at least partially, this background is formed by blazars and remnants of supernovae (SN) of type Ia in distant galaxies. However, these contributions are not sufficient to reproduce the observed level of the signal. In this work we propose another source which could contribute to this background, namely the jets of active galactic nuclei (AGN). The composition of jets is not known, but there are observational hints that the fraction of positrons there is substantial. Positrons are partially evacuated to the intergalactic medium and partially mix with the circumgalactic medium and annihilate there comparatively quickly. Using the AGN luminosity function, we estimated the positron production rate and the contribution of the positron annihilation to the cosmic background below 511~keV. We also estimated the analogous contribution from positron annihilation within SN~Ia remnants in distant galaxies. The contribution of AGNs is estimated to be a factor of 5--10 smaller than the observed background intensity, and the contribution from SNe is yet smaller by one order of magnitude. Nevertheless, the contribution of AGNs appeared to be larger than the contribution of blazars estimated from \textit{Swift}-BAT and \textit{Fermi}-LAT observations. The main uncertainty in our model is the fraction of positrons remaining in the circumgalactic medium which makes our estimation an upper limit.

\englishkeywords{active galactic nuclei, cosmic gamma-ray background}
\end{abstract}

\section{Introduction} \label{sec:intro}
Isotropic electromagnetic background in the MeV--GeV range is a manifestation of high energy
processes in the Universe. Understanding the composition of this background not only
helps to reveal the details of the physical processes at play, but is also needed for studies of 
compact sources where it is important to disentangle the contribution from the background. Over the
years of investigation, various types of objects have been proposed as sources of the
gamma-ray background, GRB from now on (see a review \cite{Fornasa2015} and Figure~9 in 
\cite{Fukazawa2022}). In \cite{Gilli2007} it was shown  from \textit{XMM-Newton} and
\textit{Chandra} observations below 10~keV and theoretical knowledge on AGN spectra in a wide energy range that
the X-ray emitting active galactic nuclei (AGN) can explain GRB up to $\sim 100$~keV.
Also, the flat spectrum radio quasars (FSRQ) observed by
\textit{Swift/}BAT can provide up to 100\% of GRB in the MeV range \cite{Ajello2009}. However,
\cite{Ajello2012} found that FSRQs observed by \textit{Fermi-}LAT can provide only
$\sim30$\% of GRB in the same range.

Another possible contributor at $\sim 1$~MeV are supernova (SN)
explosions, mainly SN~Ia. Their contribution has been estimated for decades by
different researchers despite considerable obstacles due to poor knowledge of SN~Ia rate
at large redshifts ($\gtrsim 1.5$) and differences in the models of SN explosions itself.
Among the latest works on the topic, \cite{Horiuchi2010} found that the intensity
at MeV energies from SN~Ia is almost an order of magnitude smaller than the observed
background. Later, \cite{Ruiz-Lapuente2016} carried out a new analysis and confirmed
that SNe~Ia are insufficient to account for MeV background, providing from one-third
to one-half of the GRB in this range.

GRB could be produced by cosmic rays in distant galaxies, as proposed in \cite{Lacki2014}.
The authors considered normal and star-forming galaxies and found that cosmic rays in them
could produce a large fraction of GRB at GeV energies, but still a very small fraction
at MeV energies.

In the context of AGN, it is usually assumed that the gamma-ray emission is due to
inverse Compton scattering. An alternative possibility was put forward in
\cite{Furlanetto2002}: gamma rays could stem from the annihilation of positrons contained in
the AGN jet. In \cite{Furlanetto2002} it is assumed that such positrons escape into the
intergalactic medium (IGM) and annihilate there with the electrons of the hot gas. The
authors did not consider this process as a contributor to GRB, and they only predicted
the corresponding gamma-ray flux from the Virgo cluster. In the present work, we explore
an analogous possibility, namely that the positrons contained in the AGN jet do not
escape into IGM, but remain in the gaseous halo of the host galaxy, subsequently
annihilating with its electron population. This is a continuation of our previous work
\cite{Nizamov2023} (Paper~I) where we applied the above idea to an individual galaxy M31.
Under assumption that it hosted an AGN in the past, we calculated the flux of 511~keV
photons from it, which originate from annihilation of positrons accumulated in the
gaseous halo over the lifetime of the galaxy. The resulting flux appeared to be of the order
of \textit{INTEGRAL} point source sensitivity. However, the source is expected
to span approximately $1\degree$ in the sky which could hinder the detection.
In the present work, we aim to apply the same idea to the whole population of AGN in
the Universe to estimate the potential contribution of the described process to GRB
at sub-MeV energies. In what follows, we assume $\Lambda$CDM cosmology with $\Omega_m=0.3$,
$\Omega_\Lambda=0.7$.

%%%%%%%%%%%%%%%%%%%%%%%%%%%%%%%%%%%%%%%%%%%%%%%%%%%%%%%%%%%%
\section{Positron production in individual AGN} \label{sec:1}
In calculation of the positron production in an individual AGN, we follow Paper~I. The
calculation is based on the idea that the jet is composed of protons, electrons and
positrons (and maybe ions, but we do not take them into account). Kinetic energy of
the jet is assumed to be contained in protons, as argued in \cite{Ghisellini2014}.
The main argument for this is that pure electron-positron jet should stop very quickly
due to the "Compton rocket" effect. On the other hand, there are indications that the
positrons should be present as well. For example, this could reconcile jet
energy estimations from different methods \cite{Pjanka2017}.
%%%%%%%%%%%%%%%%%%%%%%%%%%%%%%%%%%%%%%%%%%%%%%%%%%%

To derive the positron production rate, we use statistical relations given in \cite{Ghisellini2014}, namely the authors found that the total jet power is close to the total accretion power, on average: $P_\mathrm{j} = \eta \Dot{M}_\mathrm{acc} c^2$ with the coefficient $\eta \sim 1$ representing the efficiency of energy transfer from accreting material to the jet. The radiative power of the jet $P_\mathrm{rad}$ is found to be approximately an order of magnitude smaller than the total power. The authors argue that the total power is dominated by the kinetic energy of protons in the jet. They assumed that, apart from radiation and magnetic field, the jet consists of electrons and protons only (no positrons). Further, all the electrons participate in radiation and for each electron, there is a proton. When we allow for pairs in the jet so that there are  $n_\mathrm{pair}$ positrons per each proton, the kinetic power estimated from the radiation decreases by $2n_\mathrm{pair}$, because the number of protons per lepton decreases by this number, so $P_\mathrm{j} = \eta \Dot{M}_\mathrm{acc} c^2 / 2n_\mathrm{pair}$. On the other hand, the jet kinetic power equals the total energy of the protons traversing the jet cross-section per unit time: $P_\mathrm{j} = \Dot{N}_\mathrm{p} \Gamma m_\mathrm{p} c^2$ where $\Gamma$ is the jet bulk Lorentz factor. From the last two equations, noting that $\Dot{N}_+ = n_\mathrm{pair} \Dot{N}_\mathrm{p}$, we obtain
\begin{equation}
    \Dot{N}_+ = \frac{\eta \Dot{M}_\mathrm{acc}}{2 \Gamma m_\mathrm{p}}, \label{eq:ndot1}
\end{equation}
Note that the above reasoning is valid for $n_\mathrm{pair}$ smaller than 10--15, because if $n_\mathrm{pair}$ is large, the jet power is not dominated by protons anymore. However, the value of $n_\mathrm{pair}$ reported in literature is typically less than 20. Note also that the above expression allows in principle to derive the positron production rate. Combined with the total lepton content from SED fitting it could provide an estimate of the relative positron content in the jet. Of course, for such a calculation to be viable, the signal from positron annihilation should be detectable. For the jet bulk Lorentz factor we adopt the value of 10 which is typically found in statistical studies and modeling of blazars \cite{Ghisellini2015, Savolainen2010, Hervet2016}. Note also that the factor $\eta$ might be quite different for different objects, but we are interested in the contribution from the whole AGN population, and statistically this factor is close to 1 \cite{Ghisellini2014}.

In Paper~I, we tried to estimate $\Dot{N}_+$ for an individual galaxy (M31) at various
cosmic times. To do this, we had to solve the continuity equation for the supermassive black hole mass
distribution function and assign an average mass evolution to M31 because it is
impossible to reliably recover this time evolution for an individual object. In contrast, in
the present work we calculate the positron production for the whole population of
AGNs in the Universe using their luminosity function from \cite{Ueda2003}. In this case
we obviously know the luminosity of AGNs and can relate it to the mass accretion rate
via the radiative efficiency parameter $\epsilon$:
$L = \epsilon \Dot{M}_\mathrm{acc} c^2$.
We adopt $\epsilon= 0.1$ \cite{Marconi2004, Merloni2008}. Now
Eq.~\ref{eq:ndot1} reads
\begin{equation}
    \Dot{N}_+(t) = \frac{\eta \int L\phi(L, t) d\log L}{2 \epsilon c^2 \Gamma m_\mathrm{p}}, \label{eq:ndot2}
\end{equation}
where $\phi(L, t)$ is the AGN luminosity function at cosmic time $t$ and $L$ is the 
bolometric luminosity of an AGN.

To annihilate efficiently, positrons must slow down from relativistic to thermal velocities. Below 1~GeV, the main mechanism of cosmic ray positron energy loss is Coulomb collisions \cite{Prantzos2011}. However, the material of the jet is subject to adiabatic cooling which affects the positron energy much stronger than collisions. Adiabatic losses can be expressed as follows \cite{Matthews1990}:
\begin{equation}
    -\frac{d\gamma}{dt} = \frac{\gamma}{R} \frac{dR}{dt}
\end{equation}
where $\gamma$ is the particle Lorentz factor, $R$ is the source size expansion factor (i.e. $R \equiv 1$ at the start of expansion). Hence, in absence of additional energy supply, the particle energy decreases proportionally to the expansion. From observations of radio lobes of 3C~84 with VSOP \cite{Asada2006}, the age of the lobes is estimated as $\sim 45$~yr in 2001, and it is shown to have doubled its size in approximately 25 years. Numerical simulations of radio lobes show that the lobe size can increase by orders of magnitude in tens of Myr \cite{Hardcastle2018, Turner2023}. In contrast, Coulomb collisions take order of $10^9$~yr to slow down a $\sim$~GeV positron, and annihilation times are even longer (see Paper~I). Thus, we conclude that the braking time can be neglected and we assume that annihilation starts immediately when positrons are produced.

\begin{figure}
    \centering
    \includegraphics[scale=0.65]{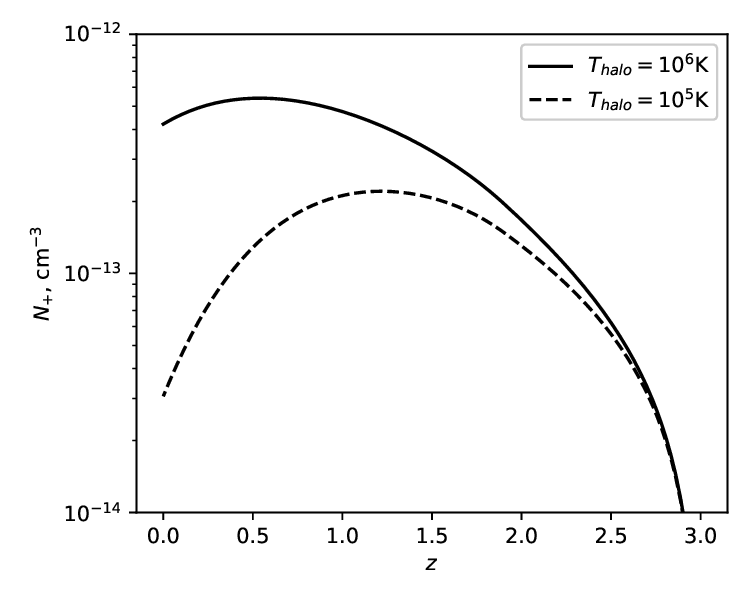}
    \caption{The mean comoving number density of positrons.}
    \label{fig:npos}
\end{figure}

The annihilation signal from each AGN depends on the amount of positrons accumulated
at a given redshift. Therefore, we need to calculate this amount from the earliest
time up to the present. We do it using the Eq.~16 from Paper~I:
\begin{equation}
    \frac{d N_+(t)}{dt}= \dot{N}_+(t-t_\mathrm{br})-n N_+(t)(\langle \sigma_\mathrm{a} v \rangle + \langle \sigma_\mathrm{r} v \rangle)
\end{equation}
where $\dot{N}_+$ is now taken from Eq.~\ref{eq:ndot1}, $n$ is the number density of  non-relativistic electrons in the halo,
$\langle \sigma_\mathrm{a} v \rangle$
and $\langle \sigma_\mathrm{r} v \rangle$ are respectively the direct annihilation rate
and the annihilation rate due to positronium formation, $t_\mathrm{br}$ is the positron braking time which we set to 0. We tried two values of the temperature, $10^6$ and $10^5$~K, and one value of the density, $10^{-4}$~cm$^{-3}$. Solving the
last equation, we obtain the comoving positron number density as a function of cosmic
time or redshift. Note that $N_+(t)$ was dimensionless in Paper~I, because it referred to
the total number of positrons in the halo. Now this quantity has a dimension of cm$^{-3}$
and refers to the mean comoving number density of positrons. The resulting $N_+(z)$ for the two
adopted values of the halo temperature is shown in
Fig.~\ref{fig:npos}.
Positrons annihilate due to
collisions with electrons of gaseous halo via two processes: direct annihilation and
positronium formation. Positronium decays into two 511~keV photons or three photons with  energies below 511~keV, the branching ratio of the two processes being 1/4 and 3/4. Therefore, the comoving production rate density of 511~keV photons is
\begin{equation}
    \varepsilon_\mathrm{2phot}(z) = 2 n N_+(z) \left(\langle \sigma_\mathrm{a} v \rangle + \frac14\langle \sigma_\mathrm{r} v \rangle\right)
\end{equation}
where $n$ is the gaseous halo density and the factor 1/4 takes into account that only
one fourth of positronium decays into two gamma rays.\footnote{Note that $n$ is the proper density while $N_+$ is the comoving density. It should be converted to the proper density via multiplication by $(1+z)^3$. However, in this case Eq.~\ref{eq:sazonov} changes as well: a factor $(1+z)^{-3}$ appears in the integral. So, the result remains the same.} Analogously, for the three photon decay, the comoving production rate density is\footnote{Although three photons are born in this process, there is no factor 3, because it is taken into account in the decay spectrum which is normalized so that $\int \tfrac{dN}{dE} dE = 3$.}
\begin{equation}
    \varepsilon_\mathrm{3phot}(z) = \frac34 n N_+(z) \langle \sigma_\mathrm{r} v \rangle
\end{equation}

%%%%%%%%%%%%%%%%%%%%%%%%%%%%%%%%%%%%%%%%%%%%%%%%%%%%%%%%
\section{Integration over AGN population} \label{sec:2}
\begin{figure}
    \centering
    \includegraphics[scale=0.7]{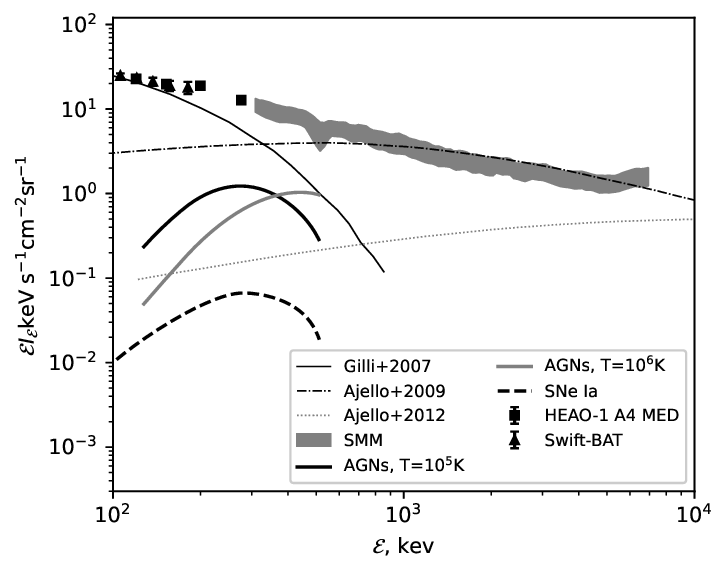}
    \caption{Diffuse GRB spectrum. The observation data are 
    HEAO-1 A4 \cite{Gruber1999}, 
    Swift-BAT \cite{Ajello2008}, SMM \cite{Watanabe1997}. The model curves represent the
    contribution from AGNs (Gilli+2007) \cite{Gilli2007} and FSRQs (Ajello+2009) \cite{Ajello2009} and (Ajello+2012) \cite{Ajello2012} .
    Our estimations from AGNs and SNe~Ia are also shown.}
    \label{fig:spec}
\end{figure}

It is possible to obtain the radiation intensity of the sources with a known redshift distribution. Let us start from Eq.~4 in \cite{Sazonov2004}:
\begin{equation}
    I_\mathcal{E} = \frac{c}{4\pi H_0}\int \frac{L[(1+z)\mathcal{E}, z]dz}{(1+z)E(z)} \label{eq:sazonov}
\end{equation}
Here $I_\mathcal{E}$ is the specific intensity in erg/cm$^2$/s/sr/erg, $\mathcal{E}$ is the photon
energy in the observer's frame, $L$ is the source luminosity density in erg/s/cm$^3$,
$E(z)=\sqrt{\Omega_m(1+z)^3+\Omega_{\Lambda}}$. In the case of two photon annihilation radiation,
the luminosity is monochromatic and equals
$L(\mathcal{E}, z) = \mathcal{E}_0\varepsilon_\mathrm{2phot}(z)\delta(\mathcal{E}-\mathcal{E}_0)$
where $\mathcal{E}_0 = 511$~keV.
Inserting this to Eq.~\ref{eq:sazonov}, we get
\begin{multline}
    I_\mathcal{E} = \frac{c\mathcal{E}_0}{4\pi H_0}
    \int \frac{\varepsilon(z)\delta[(1+z)\mathcal{E} - \mathcal{E}_0, z]dz}{(1+z)E(z)} =\\
    \frac{c\mathcal{E}_0}{4\pi H_0}
    \int \frac{\varepsilon(z)\delta[(1+z)\mathcal{E} - \mathcal{E}_0, z]d(1+z)\mathcal{E}}
    {\mathcal{E}(1+z)E(z)} =\\
    \frac{c\mathcal{E}_0}{4\pi H_0}
    \left. \frac{\varepsilon(z)}{\mathcal{E}(1+z)E(z)} \right|_{(1+z)\mathcal{E} = \mathcal{E}_0} =\\
    \frac{c\mathcal{E}_0}{4\pi H_0} 
    \frac{\varepsilon(\frac{\mathcal{E}_0}{\mathcal{E}}-1) }{\mathcal{E}\tfrac{\mathcal{E}_0}{\mathcal{E}}E\left(\tfrac{\mathcal{E}_0}{\mathcal{E}}-1\right)} = \\
    \frac{c}{4\pi H_0} \frac{\varepsilon(\frac{\mathcal{E}_0}{\mathcal{E}}-1) }{E\left(\tfrac{\mathcal{E}_0}{\mathcal{E}}-1\right)}.
\end{multline}
In the case of three photon positronium decay, the luminosity has the form $L(\mathcal{E}, z) = \mathcal{E}\varepsilon_\mathrm{3phot}(z)\varphi(\mathcal{E})$ where $\varphi(\mathcal{E})$ is the three photon decay spectrum \cite{Ore1949}.
The resulting annihilation spectrum is the sum of the two components; it is shown in Fig.~\ref{fig:spec}
in the form $\mathcal{E}I_\mathcal{E}$ with the observational data indicated in the description.
One can see that, although the possible annihilation component makes a subdominant contribution,
it can provide up to $\sim 20$\% of the total GRB in the 400--500~keV range. One should note however
that the estimate we just obtained is an upper limit because it does not take into account the
fraction of positrons which escape from the gaseous halo into IGM. We will return to this point
in Sec.~\ref{sec:discussion}.

There also could be contribution from the radiationally inefficiently accreting AGNs, which could not be accurately accounted for in our approach. Still, given that the characteristic kinetic power of the jets from these AGNs is relatively modest, $<10^{42}-10^{43}~\mathrm{erg~s}$ (see, e.g. \cite{Okamoto2008}), we would not expect that this contribution would dominate the total signal.

%%%%%%%%%%%%%%%%%%%%%%%%%%%%%%%%%%%%%%%%%%%%%%%%%%%%%
\section{Positron production in SNe~Ia} \label{sec:3}
As we mentioned in Sec.~\ref{sec:intro}, a possible source of positrons in the Universe are
explosions of SNe~Ia. We decided to make a rough estimate of their possible positron yield,
because the calculation is basically the same as we just presented. The two quantities that we
need in this case are the SN~Ia rate as a function of redshift and the average positron yield
per explosion.

The SN~Ia rate $R_\mathrm{Ia}(z)$ is usually computed as a convolution of the star formation rate (SFR) $\rho(t)$ with the
delay time distribution (DTD) $\Phi(\Delta t)$ which represents the time gap between the progenitor star formation and its explosion \cite{Horiuchi2010}:
\begin{equation}
    R_\mathrm{Ia}(z(t)) = \xi \int_{t_{10}}^t \Phi(t - t^\prime)\rho(t^\prime)dt^\prime
\end{equation}
where $\xi$ is the normalization showing the number of SNe~Ia per $M_\odot$ of stars formed, 
$t_{10}$ is the cosmic time corresponding to $z=10$, the time when the first stars were born
\cite{Horiuchi2010}. Following \cite{Chugai2023}, we adopted DTD of the form
$\Phi(\Delta t) \sim \Delta t^{-1}$ and we took the SFR fit from \cite{Yuksel2008}. Then we chose
a value for $\xi$ so that the resulting SN rate corresponded to that shown in \cite{Horiuchi2010}.

The positron yield per explosion depends on the explosion model. After \cite{Horiuchi2010}, we
assume that $0.58 M_\odot$ of $^{56}$Ni is produced in an explosion. After $^{56}$Ni decays
into $^{56}$Co*, the latter decays by electron capture or positron emission with the branching ratio
81\% and 19\%. Therefore, the positron 
production rate can be expressed as
\begin{equation}
    \Dot{n}_{+, SN} = R_\mathrm{Ia}(z(t)) \cdot 0.19 \cdot 0.58 M_\odot \frac{N_A}{M_\mathrm{Ni}}\alpha \label{eq:sne}
\end{equation}
The photon production rate for the two photon decay is simply
\begin{equation}
    \Dot{n}_{2phot, SN} = \frac12 \Dot{n}_{+, SN}\label{eq:sne2}
\end{equation}
where the factor 1/2 results from the fact that in the SN envelope (where the temperature is far below $10^6$~K) positrons annihilate mainly via positronium formation, of which only 1/4 decays into two 511~keV quanta. In Eq.~\ref{eq:sne}, the factor $\alpha$ is the fraction of positrons which eventually produce observable
gamma quanta. We will discuss it in some detail in the following section, but for now, let us put
it equal to unity to find the upper limit of the SNe~Ia contribution. $M_\mathrm{Ni}$ is the atomic mass of $^{56}$Ni. In analogy with the AGN case, the intensity for the three photon decay is obtained from Eq.~\ref{eq:sazonov} where $L(\mathcal{E}, z) = \frac34\mathcal{E}\Dot{n}_{+, SN}(z)\varphi(\mathcal{E})$. The resulting spectrum is shown in Fig.~\ref{fig:spec}. It is evident that the SN~Ia positron contribution is still smaller than that
of AGN. In fact, it is smaller than the gamma-ray lines emerging from SNe, which is evident from the comparison with \cite{Horiuchi2010}. And like the AGN contribution, this is again an upper limit because of the parameter $\alpha<1$.

%%%%%%%%%%%%%%%%%%%%%%%%%%%%%%%%%%%%%%%%%%%%%%%%%%%%%%%
\section{Discussion}\label{sec:discussion}
In the calculation of positron production in galactic halos, we took the same value of electron
temperature and density as in Paper~I, which can be regarded as typical values for a present-time
spiral galaxy. In case of cosmological sources, their evolution should be taken into account. In
\cite{Huscher2021}, the authors used the cosmological EAGLE zoom simulations to compare physical properties of gaseous halos around star forming galaxies at $z=2-3$ and $z=0$. They found that
the halos in the past and present are different in the relative proportion of cold and hot gas (i.e. gas
at the temperature less than or greater than $10^5$~K), in the metallicity and in the kinematics.
Only the density and temperature of the gas are important for our calculation of the positron annihilation rate. Data given in \cite{Huscher2021} indicate that 10~Gyr ago, i.e. at
the redshifts $z\approx 2-3$ the halos were already dense and contained both cold and hot components.
The mass of the latter corresponds to the electron number density of the order $10^{-3}$~cm$^{-3}$.
Similar results were obtained in \cite{Stern2019} using cooling flow simulations.\footnote{Note that our results hold for relatively large halo densities. We tried $n_\mathrm{halo}=2\times10^{-3}$~cm$^{-3}$ and obtained slightly lower signal: although higher annihilation rate in the dense material results in larger intrinsic luminosity, positrons most actively annihilate  in high $z$ sources the signal from which is suppressed by the huge distance.}

Direct observations of galactic halos at large redshifts are difficult and not abundant.
Observations of galaxies at $z=2-3$ in absorption lines of O, C, Si are published in
\cite{Turner2014, Rudie2019}. It is not straightforward to link them to the total halo mass.
In \cite{Rudie2019} estimates of the total C, Si and O mass are given, but one has to know individual
abundances of these elements. Using as an approximation the global metallicity of $10^{-3}$
from \cite{Huscher2021} we can derive the total halo mass and density which appears close to
$10^{-4}$~cm$^{-3}$. In \cite{Turner2014}, the total column densities are given only for
particular ions and cannot be translated to the total mass. Observations of hydrogen Ly$\alpha$
line are given but the corresponding column density is not provided because of the line saturation.

X-ray observations of the hot gas in distant galaxies are also difficult. Moreover, in the recent
paper on the Galactic halo X-ray emission observed by eROSITA \cite{Locatelli2023} it is shown
that the Milky Way's gaseous halo most probably consists of both a disk-like and a spherical
components, and the first one dominates the X-ray emission, while the second one contains the
most of the halo mass. If the same applies to other galaxies, this renders halo mass determination
in spiral galaxies from X-ray observations alone rather problematic.

X-rays from local elliptical galaxies were first reported in \cite{Forman1985}. In this work,
halos in a number of galaxies were modeled with a $\beta$-distribution with the density of
$10^{-1} - 10^{-2}$~cm$^{-3}$ in the center and few $\times 10^{-5}$~cm$^{-3}$ at 100~kpc distance,
which is close to what we suppose in this work. There is evidence that these galaxies have only
slightly evolved since $z=1-1.5$. In \cite{Civano2014} it is shown that the relation between
X-ray and optical luminosity of local elliptical galaxies established in
\cite{Boroson2011} also holds for ellipticals at $z<1.4$, once a number of factors are taken into account.\footnote{Actually, the authors found that high redshift ellipticals are more X-ray luminous,
but it can be due to a number of galaxies residing in the center of groups/clusters, or due to hidden
AGNs, or due to the slight evolution of X-ray properties of the galaxies.} In \cite{Lehmer2007} it is
shown that the X-ray to $B$-band optical luminosity ratio does not evolve strongly in the range
$z=0-0.7$, and the same conclusion is drawn for the range $z=0-1.2$ in \cite{Danielson2012}. Summing up, existence of hot gaseous halos around the early type galaxies is established with a great degree of certainty.

Another point of concern is the interaction between the AGN jet and the halo. We assumed that the
positron population of the jet remains in the halo and annihilates there eventually. The jet-halo
interaction can be studied in numerical simulations. Such a simulation was performed in
\cite{Weinberger2017}: the authors explored how much energy AGN jets can inject into the
intracluster medium. In their calculations, jet was active for 50~Myr and the calculation spanned
the time period of 336~Myr. The jet power varied from $10^{44}$ to $10^{45}$~erg/s.
It is evident from their results that the material of lower power jets ($10^{44}$~erg/s) stays
within 100~kpc from the central engine until the end of calculation. For a power of
$3 \times 10^{44}$~erg/s, only about 10\% of the injected mass remains within this radius,
and slightly less for a $10^{45}$~erg/s power jet. Another simulation \cite{Mukherjee2016}
analogously showed that low-power jets ($\lesssim 10^{43}$~erg/s) are practically trapped inside
the interstellar medium (ISM), whereas stronger jets drill through ISM relatively easily.
From AGN luminosity function of \cite{Ueda2003} it can be seen that the majority of kinetic energy
injection is due to more powerful ones, with kinetic luminosities in excess of $10^{45}$~erg/s.

Taking into account the results of \cite{Weinberger2017}, it means that most positrons produced by
AGNs are probably evacuated outside the gaseous halo into IGM. There is growing evidence for
interaction between AGN jets and interstellar or circumgalactic medium, referred to as AGN feedback,
e.g. \cite{Tumlinson2017, Donahue2022, Cielo2018, Sutherland2007}. In  the course of this feedback, material is
dragged by the jet to substantial distances, but eventually some fraction of it falls back, participating in larger baryon cycle. However, we could not find quantitative description of this process. Overall, the escape of jet material outside the circumgalactic medium makes our estimation of the GRB an upper limit.

The same is true for the estimation of the signal from SNe~Ia due to the factor $\alpha$ in
Eq.~\ref{eq:sne}. This factor takes into account that not all the photons produced in the
positron annihilation within SN envelope have chance to be observed. It should not be confused
with the positron survival fraction which is used in studies of Galactic diffuse gamma ray
emission, such as \cite{Chan1993} and \cite{Milne1999}. In those works, survival fraction is
the fraction of positrons which escape into ISM, because if a positron annihilates within the
envelope, then the corresponding photon is either not observed at all or attributed to a point
source. In other words, such a positron cannot contribute to the Galactic diffuse radiation.
In \cite{Chan1993}, the survival fraction is found to be $\sim 0.01$. In our case, the $\alpha$
parameter is larger because supernovae
reside in distant galaxies, so we only demand that the positron annihilates in an optically
thin medium, probably within the envelope. Let us estimate the optical depth of the
envelope for some typical parameters of its mass $M=1M_\odot$ and velocity $v=10^4$~km/s.
It takes 120 days for $^{56}$Ni to decay into $^{56}$Co and then to $^{56}$Fe \cite{Martin2010}.
With known mass, expansion velocity and time and the attenuation factor
$8.63 \times 10^{-2}$~cm$^2$/g \cite{Martin2010} we obtain the optical depth of the envelope
$\tau=0.13$ and $\alpha=e^{-\tau}=0.88$. Note however that due to this exponential dependency,
$\alpha$ is sensitive to the explosion parameters. If ejecta mass is larger and velocity smaller,
it would decrease  $\alpha$. Moreover, the velocity is not uniform throughout the ejecta, but rather
grows from the inner edge to the outer. A more thorough estimation of $\alpha$ requires modelling
of the explosion which is beyond the scope of this work.

Our results could be compared to other models in this energy range, as one can see from
Fig.~\ref{fig:spec}. The MeV-range background estimated from \textit{Fermi} observations
of FSRQs \cite{Ajello2012} appears to be lower than our AGN estimate, but
higher than our SN~Ia estimate. On the other hand, the hard X-ray background originating
from Compton-thin and Compton-thick AGN computed in \cite{Gilli2007} and the background modeled in
\cite{Ajello2009} on the basis of \textit{Swift} observations of FSRQs are above all our estimations.
Note however that in \cite{Ajello2009} observations in the 15--55~keV band were extrapolated to
higher energies and only 18 FSRQs were used for the background estimation. In \cite{Ajello2012}
observations of both \textit{Swift}-BAT and \textit{Fermi}-LAT were used to infer SED and the
sample comprised of 186 FSRQs. 
It can be seen from Fig.~\ref{fig:spec} that the observed background intensity in the range 
$\sim 300-500$~keV is substantially larger than the current models, including the one considered
in this work. However, the mechanism we proposed can contribute a non-negligible part of this
diffuse radiation.

%%%%%%%%%%%%%%%%%%%%%%%%%%%%%%%%%%%%%%%%%%%%%%%%%%%%%%%%%%%%%%%%%%%%%%%%%%%%%%%%%%%%%%%%%%%%%
\section{Conclusions}
In this paper we estimated the contribution to the diffuse gamma ray background which can be made
by cosmological sources of positrons via subsequent annihilation of these particles. We considered
two types of sources. First, positrons from AGN jets which annihilate in gaseous halos around
AGN host galaxies. Second, SNe~Ia in distant galaxies which mostly annihilate in the SN ejecta. A narrow line near 511~keV due to two photon annihilation is smeared because of cosmological expansion. Moreover, part of the signal is due to three photon annihilation which has a continuous spectrum below 511~keV.
Our calculations show
that the signal due to AGN jets is approximately an order of magnitude below the observed intensity
at 511~keV and even smaller at lower energies. The contribution from SNe~Ia is approximately two
orders of magnitude below the observed emission level. Both estimates are upper limits. In the case
of AGNs, we do not know the fraction of positrons which remain in the halo, while in the case of
SNe we do not know the fraction of positrons which annihilate in sufficiently transparent medium.
In the latter case this proves that the contribution from SNe~Ia cannot be a substantial part of
the diffuse GRB. In fact, this has been shown in previous works, such as \cite{Horiuchi2010}. In
this work the contribution from SNe~Ia was estimated to be at most 10--20\%, but the calculation
included nuclear gamma-ray lines. From our calculation, it is evident, that the contribution from
SN positrons is subdominant when compared to that from SN nuclear lines.

The contribution from AGN jet positrons seems more promising. in the range 200-500~keV, it is
below the estimate of \cite{Ajello2009}, but above the estimate of \cite{Ajello2012} which,
we believe, is more reliable due to larger statistics and use of  two instruments,
\textit{Swift}-BAT and \textit{Fermi}-LAT. We emphasize once again that our estimation is
an upper limit, but still the mechanism itself could contribute a considerable fraction of diffuse
GRB in the sub-MeV range.

\section*{Acknowledgements}
The work of the authors was supported by the Ministry of Science and Higher Education of Russian Federation
under the contract 075-15-2020-778 in the framework of the Large Scientific Projects program within the national
project "Science". 

\bibliographystyle{plainnat}
\bibliography{main}

\end{document}